\documentclass[nocite,overfull]{epl}

\title{Entropic Elasticity of Phantom Percolation Networks}
\shorttitle{Entropic Elasticity of Phantom etc.}
\author{O. Farago \and Y. Kantor}
\institute{School of Physics and Astronomy, Raymond and Beverly
  Sackler Faculty of Exact Sciences, Tel Aviv University, Tel
  Aviv 69 978,  Israel}

\begin{document}

\maketitle

\begin{abstract}
A new method is used to measure the stress and elastic constants of
purely entropic phantom networks, in which a fraction $p$ of neighbors
are tethered by inextensible bonds. We find that close to the
percolation threshold $p_c$ the shear modulus behaves as $(p-p_c)^f$,
where the exponent $f\approx1.35$ in two dimensions, and $f\approx1.95$
in three dimensions, close to the corresponding values of the
conductivity exponent in random resistor networks. The components of
the stiffness tensor (elastic constants) of the spanning cluster follow
a power law $\sim(p-p_c)^g$, with an exponent $g\approx 2.0$ and 2.6
in two and three dimensions, respectively.  
\end{abstract}

In the gelation process, monomers or short polymers in a fluid
solution are randomly cross-linked. At a certain moment during the
reaction, a macroscopically large network, {\em the gel}\/, spans the 
system. At this point, the system changes from a fluid-like (sol) to a
solid-like (gel) phase that has a finite shear modulus. The {\em
  geometry}\/ of gels is frequently described by the percolation model
\cite{stauffer}. The percolation geometry is usually defined on a
lattice, by randomly occupying a fraction $p$ of the bonds (or
sites). The gel point is identified with the percolation threshold
$p_c$, the critical bond (site) concentration above which a spanning
cluster is formed. Percolation theory predicts that close to $p_c$
quantities like the average cluster size or the gel fraction have
power laws dependence on $(p-p_c)$ with universal exponents, some of
which have been measured experimentally for gel systems \cite{adam}.  

Near the sol-gel transition typical polymer clusters are very large,
tenuous and floppy. Elastic properties of such systems are primarily
determined by the {\em entropy}, i.e., distortions of a sample barely
modify its energy, but they decrease the available phase space
(decrease entropy) and, thus, increase the free energy. Like
geometrical quantities near $p_c$, the shear modulus is also expected
to follow a power law: $\mu\sim(p-p_c)^f$. De Gennes
\cite{degennes} used an analogy between gel elasticity and
conductivity of random resistor networks (RRN), and conjectured that
the exponent $f$ should be equal to the exponent $t$ describing the
conductivity $\Sigma$ of RRN close to $p_c$:
$\Sigma\sim(p-p_c)^t$. Alternative theories take different approaches
and lead to different exponents \cite{other}. An exact calculation of
the critical behavior of $\mu$, which takes into account excluded
volume (EV) and entanglements effects, is not yet
available. Experimental values of $f$, measured for different
polymeric systems, are very scattered \cite{experiments}. One of the
reasons for the variety of the experimental results is the mixing of
the entropic and energetic contributions to the gel elasticity, which
influences the ``effective'' exponent.

Neglect of EV interactions, i.e., treating a {\em phantom}\/ system,
may strongly modify the physics. Nevertheless, it is frequently done
either because in certain situations (such as dense polymer melts) EV
interactions effectively cancel out \cite{dgbook}, or because from
purely theoretical point of view phantom systems are more tractable
and may serve as a starting point for studying real systems. Phantom
systems maintain the correct connectivity, which is one of the
important characteristics of a polymer network. A feature common to
most phantom networks (independently of the detailed shape of the
microscopic potential) is the fact at zero tension the probability
density that two distant nodes are separated by $\vec{r}$, takes a
Gaussian form $\sim\exp\left[-\frac{1}{2}Br^2\right]$. For linear
polymers this is a consequence of the central limit theorem, while for
more complicated systems this can be demonstrated numerically
\cite{kantor}. The thermodynamic behavior of a phantom network can,
therefore, be described very accurately within a phantom Gaussian
network (PGN) model, in which each bond of the network is
replaced by a {\em Gaussian spring}\/ having the energy
$E=\frac{1}{2}Kr^2$, where $r$ is its end-to-end distance
\cite{fisherman}. Corrections to Gaussian behavior can be studied by
considering a phantom nearly-Gaussian network (PNGN), in which the
springs' energies include an additional small term equal to
$\frac{1}{4}ar^4$. In this paper we describe results of a numerical
study of the elasticity of tethered phantom percolating networks, and
compare our results with the two models \cite{gaussian}. We use a
recently developed formalism \cite{fluctuation} which enables direct
calculation of entropic elastic constants of tethered systems. We show
that the shear modulus behaves near $p_c$ like the conductivity of RRN
as predicted by the PGN model, while the elastic stiffness tensor of
the spanning cluster, which according to PGN model is supposed to
vanish \cite{gaussian}, also exhibits a power law behavior near $p_c$
with a significantly larger critical exponent. The last result is a
consequence of the deviation from the Gaussian behavior, and can be
understood within the PNGN model \cite{gaussian}.   

In a homogeneous deformation we distort the boundaries of a system is
such a way that a separation $\vec{R}$ between a pair of surface
points is modified into $\vec{r}$, which is linearly related to
$\vec{R}$ via a position-independent matrix. In this case the new
squared distance $r^2=R_iR_j(\delta_{ij}+2\eta_{ij})$, where the
subscripts denote Cartesian coordinates, $\delta_{ij}$ is the
Kr\"{o}necker delta, $\eta_{ij}$ is the Lagrangian strain tensor, and
summation over repeated indices is implied. The stress and elastic
stiffness tensors, $\sigma_{ij}$ and $C_{ijkl}$, respectively, are then
defined as the coefficients of the expansion of the free energy
density in the strain variables:
$f(\{\eta\})=f(\{0\})+\sigma_{ij}\eta_{ij}+\frac{1}{2}C_{ijkl}\eta_{ij}
\eta_{kl}+\ldots$\, Close to $p_c$, percolation networks
``forget'' the details of the lattice and behave like isotropic
systems, and therefore the stress tensor
$\sigma_{xx}=\sigma_{yy}=\sigma_{zz}\equiv\sigma\equiv-P$, where $P$ 
is the pressure. Isotropic systems have only three {\em different}\/
non-vanishing elastic constants: $C_{11}\equiv
C_{xxxx}=C_{yyyy}=C_{zzzz}\,$; $C_{12}\equiv
C_{xxyy}=C_{yyzz}=C_{zzxx}=\ldots\,$; and
$C_{44}\equiv\frac{1}{2}(C_{xyxy}+C_{xyyx})=\frac{1}{2}(C_{yzyz}
+C_{yzzy})=\ldots\,$, which are related by:
$C_{11}=C_{12}+2C_{44}$. Frequently, one finds it more useful to
describe the elastic behavior in terms of the {\em shear}\/ modulus   
$\mu\equiv C_{44}-P$, and the {\em bulk}\/ modulus $\kappa\equiv
\frac{1}{2}(C_{11}+C_{12})$ [for two-dimensional (2D) systems], or
$\kappa\equiv\frac{1}{3}(C_{11}+2C_{12}+P)$ [for three-dimensional
(3D) systems]. In a percolation phantom system the contribution
of the different clusters are additive. Each finite cluster, not
connected to the boundaries of the system, contributes as a single
atom of an ideal gas. Thus $N_0$ free finite clusters confined within
volume $V$ at temperature $T$, produce stress equal to
$\sigma=-\frac{N_0kT}{V}$, and elastic constant
$C_{44}=\frac{N_0kT}{V}$, where $k$ is the Boltzmann
constant. Although both $C_{44}$ and $P$ are affected by the presence
of finite clusters, we observe that the (ideal gas) contribution
of the finite clusters cancels out in the definition definition of
$\mu$. Since finite clusters play such unremarkable role in the
problem of elasticity, we will disregard them completely, and in the
remainder of this work the stress, elastic constants and elastic
moduli will refer to the contribution of the spanning cluster alone. 
The latter depends on the details of the potential and
connectivity. However, in the case of PGN few simple properties exist
\cite{gaussian}: (1) Since the energy of a Gaussian spring is
proportional to $r^2$, while the squared distances between the points
on the boundaries are linear in $\eta_{ij}$, it can be shown the free
energy of the spanning cluster does not include quadratic terms in
$\eta_{ij}$, and therefore its $C_{ijkl}=0$. (2) The stress tensor is
{\em equal}\/ to the conductivity tensor of an equivalent resistor
network in which each spring of the spanning cluster with a force
constant $K$ is replaced by a resistor of conductance~$K$. For
isotropic PGNs $\mu=\sigma=\Sigma$, where $\Sigma$ is the conductivity
of the equivalent resistor network, and we, thus, find that
$f=t$. In the PNGN model near $p_c$, $\sigma$ and $\mu$ are still
dominated by the Gaussian term rather than by the non-Gaussian
perturbation, and we recover the equality $f=t$. Non-Gaussian
corrections are manifested by non-vanishing elastic constants, which
for percolation PNGNs are expected to behave as $C\sim(p-p_c)^g$ with
$g>f$~\cite{gaussian}. 

\begin{figure}
\twoimages[scale=0.55]{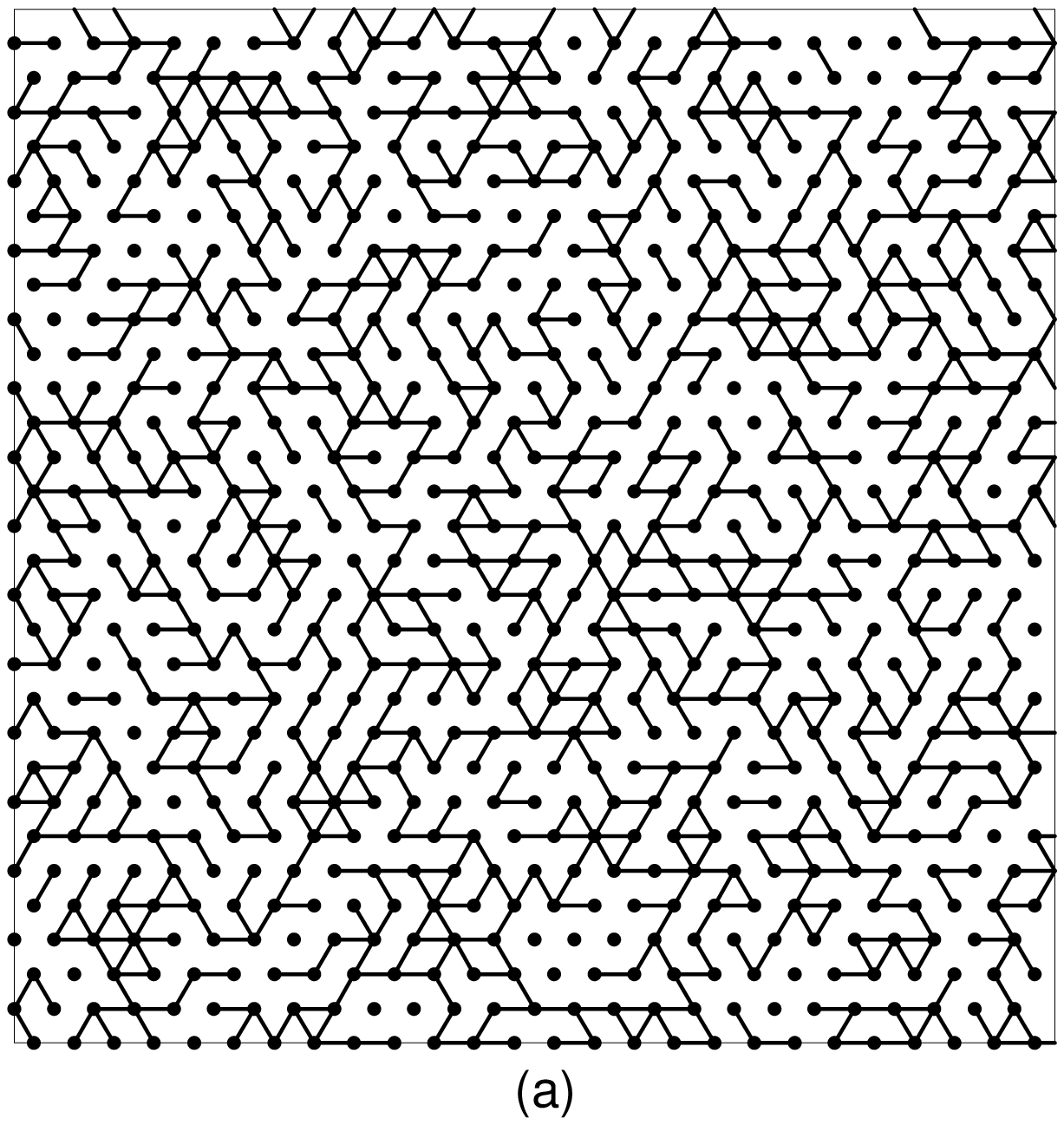}{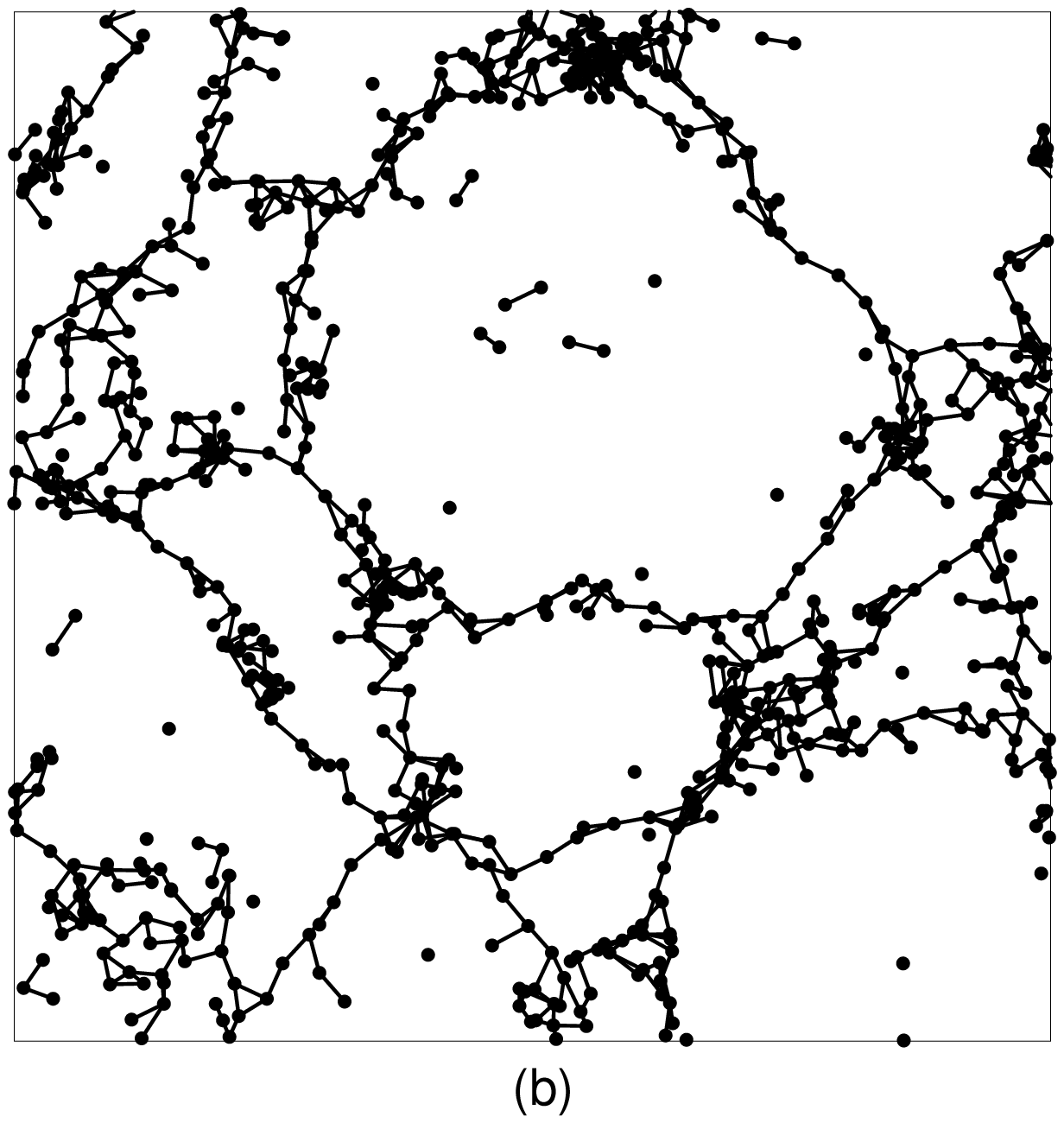}
\caption{Part of initial (a) and equilibrated (b) configurations
  of the 2D system ($p=0.405$, $b=1.05$).}
\label{config}
\end{figure}

In this work we investigate the elastic behavior of networks for which
the problem of mixing of the entropic and energetic components does
not exist. Our systems consists of point-like atoms connected by
``tethers'' that have no energy, but simply limit the distance of a
connected pair to be smaller than some value $b$. Since the internal
(potential) energy of the system vanishes, its thermodynamic behavior
is {\em purely}\/ entropic. We generated the (quenched) topologies by
considering bond percolation problem on 2D triangular
($p_c=\frac{\pi}{9}\sim  0.349$) and 3D faced-centered-cubic
($p_c\simeq0.12$) lattices, with a fraction $p$ of bond present. Each
present bond was replaced by a tether, while each site became an
``atom'' without EV, and the system was allowed to move in {\em
  continuum}. Fig.~\ref{config} (a) depicts an initial 2D
configuration of the system, which equilibrates into configuration of
the kind depicted in fig.~\ref{config} (b). As expected, finite
clusters and dangling ends of the spanning cluster contract relative
to their linear size in the initial quenched construction
\cite{cates}. The size of the backbone, on the other hand, is fixed by
the boundary conditions and, therefore, it looks like a collection of
loops of the size of the percolation correlation length.    

\begin{figure}
\twofigures[scale=0.45]{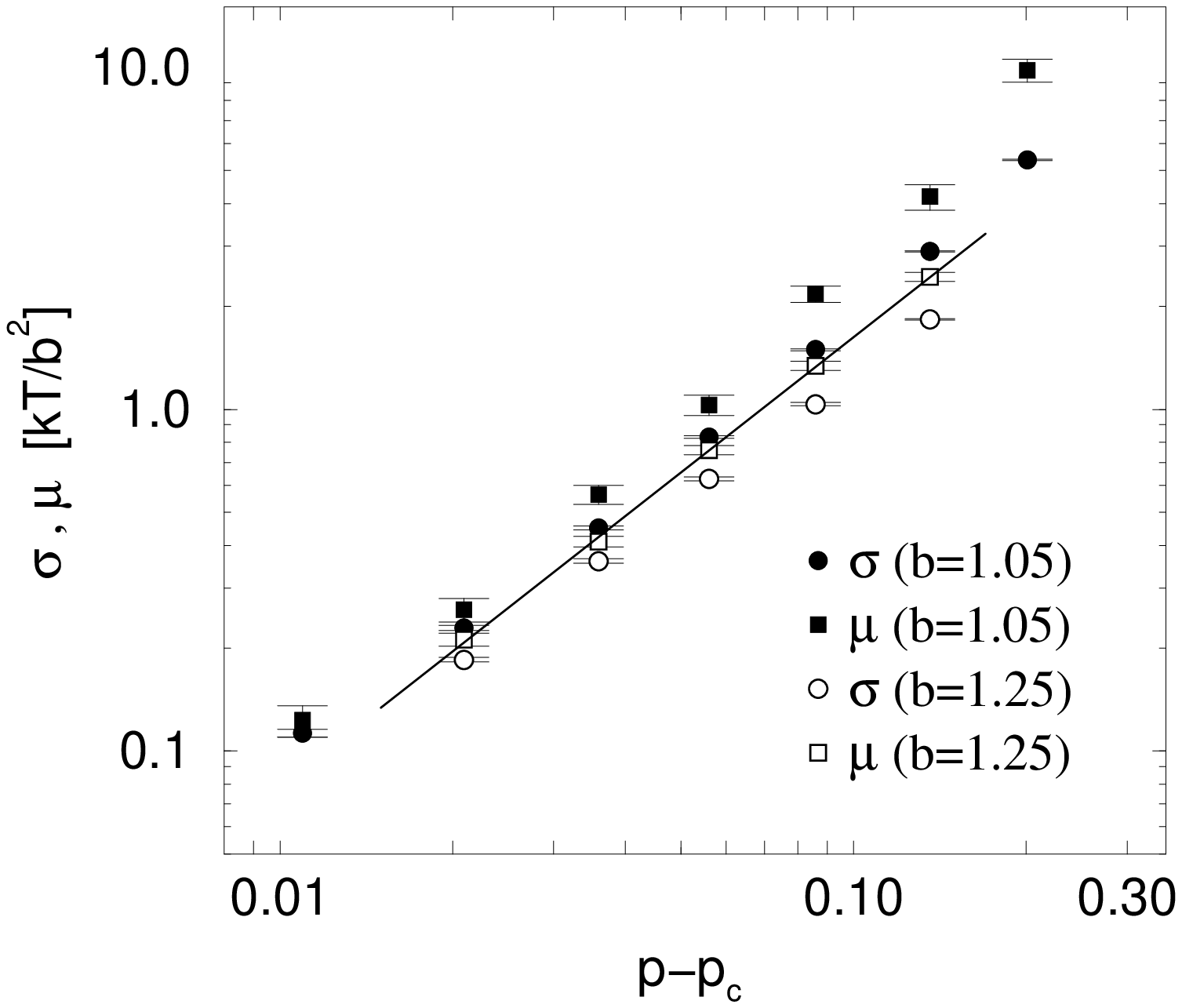}{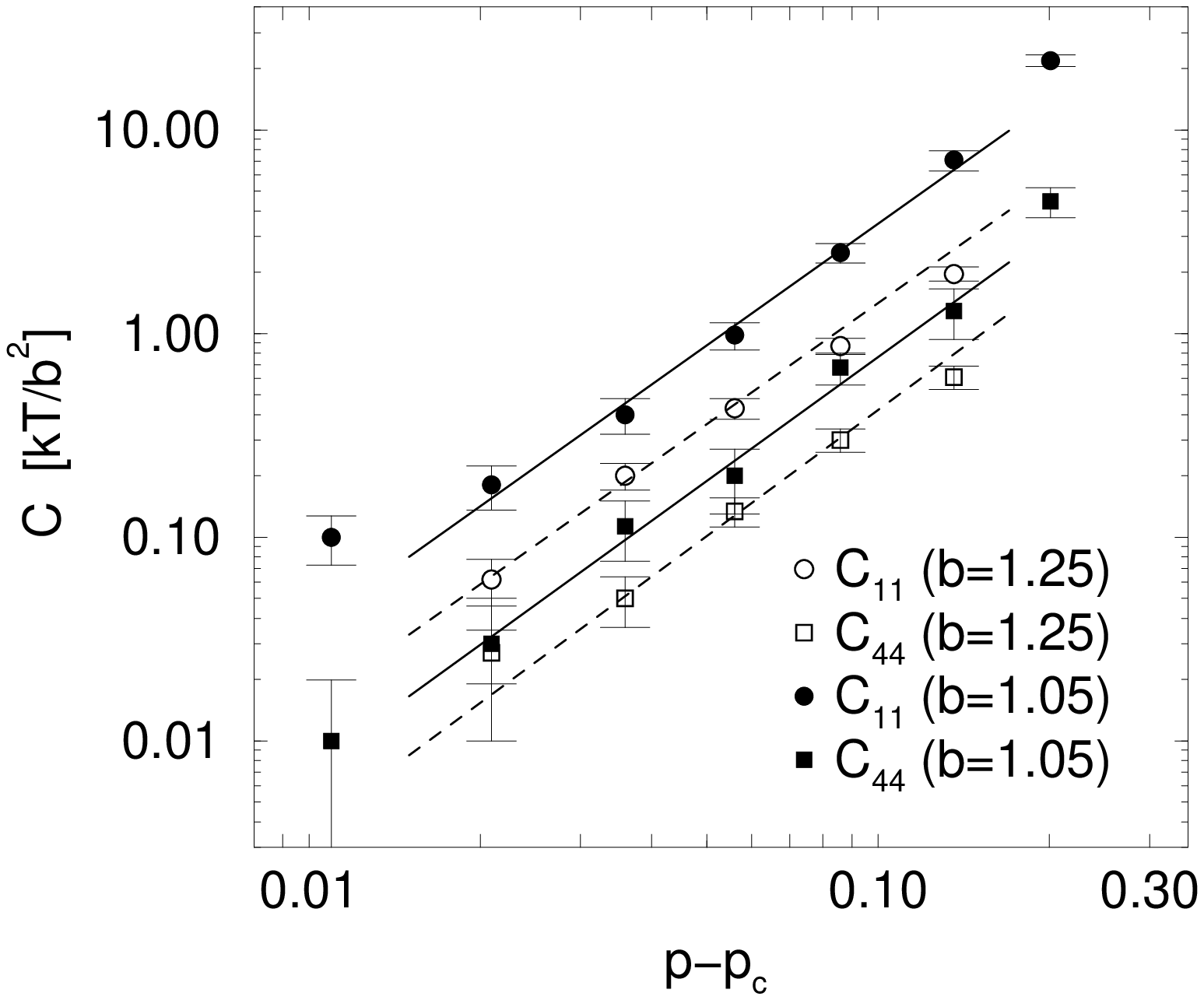}
\caption{Logarithmic plot of the stress $\sigma$ and the shear
  modulus $\mu$ as a  function of $(p-p_c)$, for 2D systems. 
  The slope of the solid line is 1.35. 
  Results are in $kT/b^2$ units.} 
\label{pmu2d}
\caption{Logarithmic plot of the elastic constants $C_{11}$ and
  $C_{44}$ as a function of $(p-p_c)$, for 2D systems. 
  The slope of the lines is $\approx 2$. 
  Results are in $kT/b^2$ units.}  
\label{c2d}
\end{figure}
The non-interacting character of phantom networks significantly
simplifies the numerical procedure: (1) Since we are not interested in
the (trivial) contribution of the finite clusters they were removed
from the simulations. (2) Dangling ends of the spanning cluster do not
contribute neither to the stress nor to the elastic constants and,
therefore, they can also be removed. Thus, for every quench we
identified the backbone (using the ``burning'' algorithm
\cite{burning}, which was slightly modified to deal with the periodic
boundary conditions applied in the simulations), and explored the
configuration space using a Monte Carlo (MC) updating scheme
\cite{jaster} in which the conventional Metropolis single atom steps
are replaced by collective steps of chains of atoms. At each MC time
unit we made a number of move attempts (with acceptance probability
$\sim 0.5$) equal to the number of atoms. In the 2D simulations, we
used a $120\times 138$ triangular lattice (that has an aspect
ratio very close to 1) with nearest-neighbor spacing $b_0\equiv 1$,
and a number of quenched topologies that ranged from $N_{t}=200$ for
$p$ closest to $p_c$, down to $N_t=20$ far from $p_c$. In the 3D
simulations we used systems of $24^3$ cubic unit cells (each
containing 4 atoms), i.e., of linear size $L=24\sqrt{2}b_0$, with
nearest-neighbor spacing $b_0\equiv 1$, and $30\leq N_{t}\leq150$. The
duration of the MC run of each individual sample was {\em at least}\/
50 times larger than the relaxation time which we estimated from the
expression $\tau=dkTL^2\rho/(\pi^2\mu s^2)$, where $s$ is the
(average) distance an atom moves in one MC time unit, $\rho$ is the
number density of atoms, and $d$ is the dimensionality of the system
\cite{relax}. The value of $\mu$ in this expression was taken, a
posteriori, from the simulations. We used a new method enabling the
direct measurement of the stress and elastic constants from the
probability densities of finding maximally extended tethers
\cite{fluctuation}. The error estimates are affected by the
fluctuations in the values of the measured quantities between the
different quenches, and to lesser extent by the thermal uncertainties
within each sample. The error bars appearing in the graphs correspond
to one standard deviation of the average. 

In our system we can vary only two non-trivial parameters: bond
concentration $p$, and the maximal tether length $b$ (measured in the
units of the nearest-neighbor spacing $b_0$). Fig.~\ref{pmu2d} depicts
our results for $\sigma$ and $\mu$ as a function of $(p-p_c)$ for 2D
systems with $b=1.05$ and $b=1.25$. It clearly demonstrates that close
to $p_c$, the network becomes Gaussian: First, the difference between
$\mu$ and $\sigma$ decreases as we approach $p_c$, which implies that
the elastic constant $C_{44}=\mu-\sigma$ vanishes faster than both
quantities. Second, when plotted in $kT/b^2$ units, the values of
$\sigma$ and $\mu$ in systems with different $b$ converge towards
each other. This is explained by the facts that (a) the stress of a
2D PGN depends only on the topology of the network and the value of
the springs force constant $K$; and (b) for the tethered networks, the
{\em effective}\/ $K$ is proportional to $kT/b^2$
\cite{fisherman}. Third, the value of $f$ extracted from the the
graphs is $f=1.35\pm0.10$, very close to the value of the conductivity
exponent $t=1.297\pm0.007$ in 2D \cite{conductivity2d}. Similar result
for the exponent $f$ has been obtained by Plischke 
{\em et al.}~\cite{plischke}. They used central force networks in 
which both entropy and energy contribute to the elastic properties
and,  by examining systems at several temperatures, removed the
energetic component. Close to $p_c$ elasticity of central force
systems is completely dominated by entropy, and their result for $f$
reflects this fact.  

In fig.~\ref{c2d} we present our results for the elastic constants
$C_{11}$ and $C_{44}$, which are supposed to vanish in the purely
Gaussian case. Shorter tethers correspond to larger values of the
elastic constants, since they represent more stretched networks, which
exhibit stronger deviations from Gaussian behavior. Despite almost an
order-of-magnitude difference between $C_{11}$ and $C_{44}$ for the
same $b$, and half an order-of-magnitude difference between the same
constants for the different values of $b$, all the results can be
described by a power law $(p-p_c)^g$, with the same exponent
$g=2.0\pm0.2$, which is significantly larger than $f$. (We do not show
the elastic constant $C_{12}$, which has large statistical
uncertainties that prevent exact determination of the power law. The
results are, however, consistent with the power laws for the other
constants.) To further ascertain the universality of $g$, one would
need to increase $b$ to even larger values. This, however, would
further decrease the values of the elastic constants which in our
method of simulations \cite{fluctuation} would increase the
statistical uncertainties, and require increase of the simulation
length beyond our computational ability.  

\begin{figure}
\twofigures[scale=0.45]{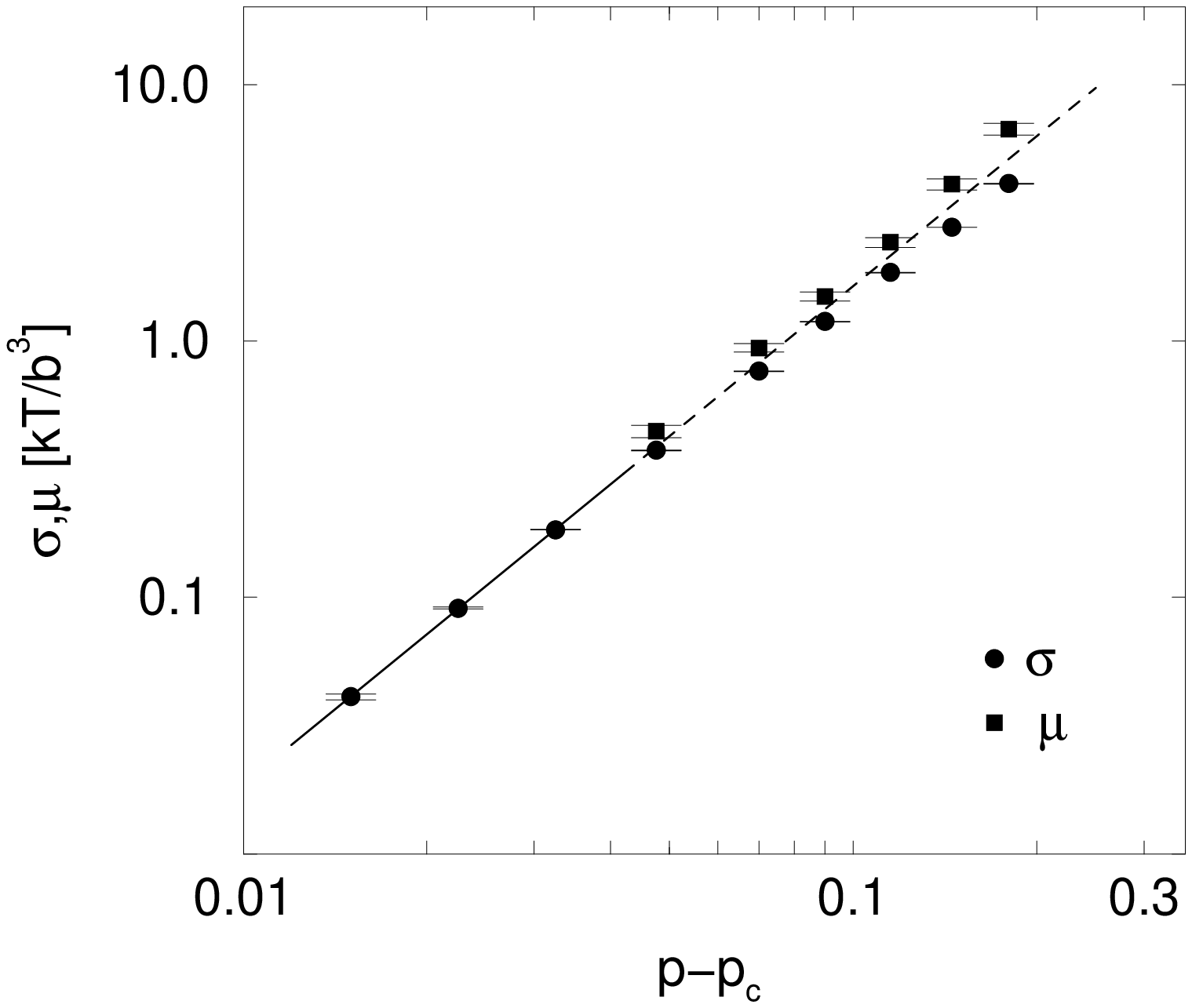}{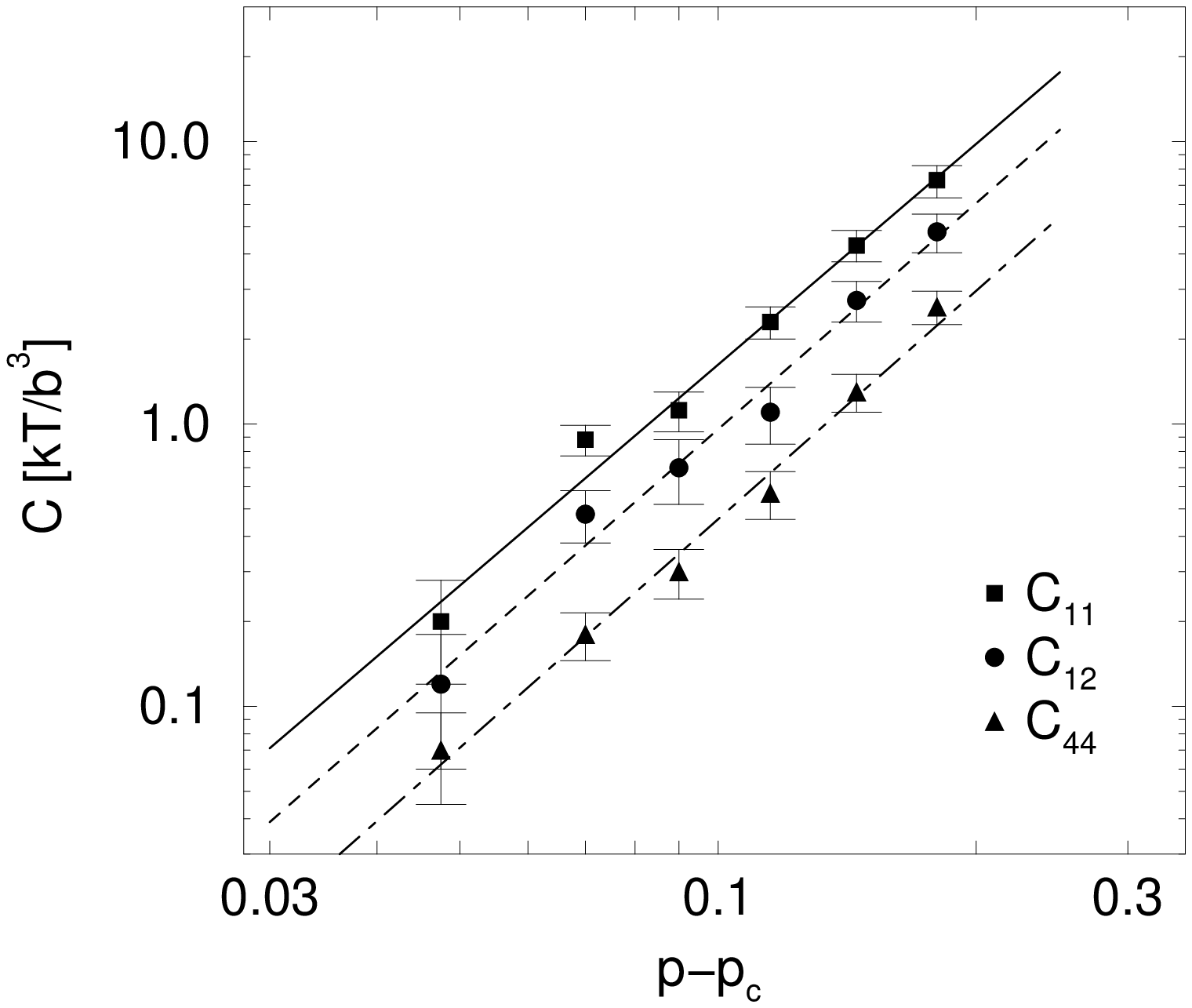}
\caption{Logarithmic plot of the stress $\sigma$ and the
  shear modulus $\mu$ as a function of $(p-p_c)$, for 3D 
  systems with $b=1.05$. The slope of the solid line is 1.95. Results
  are in $kT/b^3$ units.}  
\label{3dpmu}
\caption{Logarithmic plot of the elastic constants $C_{11}$,
  $C_{12}$, and $C_{44}$ as a function of
  $(p-p_c)$, for 3D systems with $b=1.05$. The slope of different
  lines is $\approx 2.65$. Results are in $kT/b^3$ units.} 
\label{c3d}
\end{figure}
Our results for the 3D networks with $b=1.05$ are shown in figures
\ref{3dpmu} and \ref{c3d}. Again, the validity of the PGN model is
supported by the observation that $\sigma$ and $\mu$ converge
towards each other as we approach $p_c$, following power laws
with $f=1.95\pm 0.05$, which agrees with the conductivity exponent
$t=2.003\pm0.047$ in 3D \cite{conductivity3d}. The elastic constants
also follow power laws with an exponent $g=2.65\pm0.15$. Note that our
results confirm the relation $C_{11}=C_{12}+2C_{44}$, which indicates
that close to $p_c$, percolating networks behave as isotropic
systems. At $p=1$ the system has a lower (cubic) symmetry, and there
is a gradual deviation from this relation with increasing $p$ beyond
the regime shown in fig.~\ref{c3d}. 

In ref.~\cite{gaussian} we used PNGN model to derive bounds on the
exponent $g$, which is a consequence of non-Gaussian behavior:
$3t-2\nu(d-1)\leq g\leq4(t-1)-\nu(3d-4)$, where $\nu$ is the
correlation length exponent. The perturbative derivation of the bounds
was self-consistent only for $d\geq3$. We note that our result for $g$
in $d=3$ is, within the statistical uncertainty, consistent with the
bounds $2.48\leq g\leq 2.60$. One should keep in mind, however, that
the PNGN model assumes that the coefficient of the quartic
perturbation term to the Gaussian spring energy is a constant number,
while for the tethered network model its effective value may depend on
the mean stress and, thus, on the position of the bond in the network.

In conclusion, we studied the critical elastic behavior of purely
entropic phantom model with topology of a percolating network. The
microscopic tethering potential is very different from a Gaussian
spring. Nevertheless, diluted networks become very ``floppy'' so that
the potentials become effectively Gaussian and, consequently, the
shear modulus behaves as the conductivity of RRN. Non-trivial power
law dependence of the elastic constants of the spanning cluster on
$(p-p_c)$ is a signature of a deviation from the Gaussian behavior,
and is controlled by a critical exponent significantly larger than the
exponent of conductivity. Since $g$ characterizes a ``sub-leading''
behavior, a detailed study of a broad class of potentials is needed to
verify its universality. 

\acknowledgments
We greatly appreciate the support of E.~Bielopolski and Ch.~Sofer 
from the Tel Aviv University Computation Center, and of G.~Koren from
the High Performance Computing Unit at the Inter University
Computation Center. This work was supported by the Israel Science
Foundation through Grant No. 177/99.


\begin{thebibliography}{0}
\bibitem{relax}
This expression is a generalization of Eq.(4.7) in Ref.{\protect
  \cite{kantor}}.
\end{thebibliography}
\end{document}